# Design and Evaluation of an AI-Driven Personalized Mobile App to Provide Multifaceted Health Support for Type 2 Diabetes Patients in China

YIBO MENG
mengyb22@tsinghua.org.cn

Zhingming Liu
1025189065@qq.com

Xiaochen Qin
2322269425@qq.com

*Abstract*—Type 2 diabetes patients in China face many significant challenges in patient-provider communication and self- management. In light of this, this work designed, implemented, and evaluated an AI-driven, personalized, multi-functional mobile app system named T2MD Health. The app integrates real-time patient- provider conversation transcription, medical terminology interpretation, daily health tracking, and a data-driven feedback loop. We conducted qualitative interviews with 40 participants to study key user needs before system development and a mixed- method controlled experiment with 60 participants after to evaluate the effectiveness and usability of the app. Evaluation results showed that the app was effective in improving patient-provider communication efficiency, patient understanding and knowledge retention, and patient self-management. Patient feedback also revealed that the app has the potential to address the urban-rural gap in the access to medical consultation services to some extent. Findings of this study could inform future studies that seek to utilize mobile apps and artificial intelligence to support patients with chronic diseases.

## I. INTRODUCTION

Type 2 diabetes mellitus (T2DM) has become one of the world's most persistent health challenges, now affecting more than 460 million people and continuing to rise each year [1]. China, with the largest population, accounts for an estimated 109.6 million adults with diabetes as of 2017 [2]. Managing T2DM remains difficult for many patients, particularly in rural areas, where limited medical resources, fragmented follow-up, and low health literacy are common [3–6]. Beyond these structural barriers, social and cultural factors such as dialect differences, disease stigma, and high carbohydrate dietary customs further complicate daily self-care [7–10]. Conventional clinical workflows seldom provide the time or continuity required to resolve these issues [11].

These challenges call for new approaches that extend beyond traditional clinical care. Recent advancements in artificial intelligence (AI) have created opportunities to enhance patient-provider communication, improve continuity of care, and support patient self-management [12]. Progress in speech recognition, natural language processing, and large language models has enabled real-time transcription, context-aware explanation, and personalized health dialogue generation [25, 26]. These persistent challenges call for new approaches that extend beyond traditional clinical care. Recent advancements in artificial intelligence (AI) have introduced new opportunities for medical applications that support both patients and providers [12]. Advances in speech recognition, natural language processing, and large language models have enabled real-time transcription, context-aware explanation, and personalized health dialogue generation, creating new opportunities for patient-centered digital health systems [25,26].

Building on these developments, we developed **T2MD Health**, an AI-driven, multifunctional mobile application designed to assist people living with T2DM. The system aims to (1) enhance doctor–patient communication through real-time transcription and dialect-adapted interpretation [13]; (2) support continuous care through conversational AI and structured data feedback [14, 15]; (3) improve the efficiency of information flow between patients and physicians [16]; and (4) reduce the urban–rural healthcare gap by providing personalized guidance and lowering consultation costs [17, 18]. The app integrates speech recognition, a medical knowledge graph for term localization, and GPT-4o–based adaptive education within a multimodal interface that supports both clinical consultations and self-management.

While many previous AI applications in diabetes management have focused on glucose prediction, remote monitoring, or chatbot-based education [26], few have addressed the communication and linguistic barriers that shape patient experience during clinical encounters. T2MD Health bridges this gap by combining intelligent transcription, context-aware explanation, and personalized education, fostering more natural interaction and mutual understanding between patients and providers. Beyond technical innovation, this approach demonstrates how culturally and linguistically informed AI design can contribute to equitable digital health in resource-limited settings.

We conducted qualitative interviews to identify user needs and pain points before system development, followed by a controlled experiment and post-use interviews to evaluate usability and preliminary effectiveness. The study design and evaluation methods are detailed in the following sections.

## II. USER NEEDS ASSESSMENT

### A. Study Design

We first conducted semi-structured interviews with 40 participants, including 36 patients with type 2 diabetes mellitus (T2DM) and 4 physicians, to explore the challenges they faced in T2DM treatment and management. Participants were recruited online via RedNote between November 2024 and December 2024, ensuring a balanced geographical and urban–rural distribution. Each interview lasted approximately 35 minutes and covered topics such as patient–provider communication, treatment and rehabilitation experiences, and other difficulties encountered in daily life.

## B. Interview Findings

The 36 patient participants had a mean age of XX years (SD = 14.0) and an average disease duration of 8.2 years (SD = 5.5). Among them, 56% (n = 20) were female, and 47% (n = 17) lived in rural areas. The four physician participants had an average age of 46.2 years; two were female, and three practiced in rural settings. The interviews revealed several major themes concerning difficulties in T2DM treatment and management.

### 1) Poor patient-provider communication

Participants reported significant challenges in communicating effectively with healthcare providers across all stages of T2DM care, including consultation, treatment, and rehabilitation. Many patients found it difficult to understand medical terminology, the causes of T2DM, their own health conditions, and treatment plans provided by physicians.

For instance, Participant 1 (P1) shared, *"The doctor did not explain it clearly, and I still don't understand why diabetes patients cannot eat too much flour and rice. Why can't we eat them if they are not sweet?"* — reflecting limited understanding of the concept of sugar and the etiology of T2DM. Participant 4 (P4) added, *"Why did the doctor tell me to drink more water and eat more vegetables? Drinking more water will increase urination."* This illustrates a misunderstanding of the physician's treatment advice. Such communication gaps, which ideally should be resolved during consultations, often remain unaddressed.

### 2) Limited access to medical resources

According to participants, medical resources in China are becoming increasingly scarce, particularly in rural regions. Township health centers, the primary institutions within the country's three-tier rural healthcare network, often lack qualified endocrinologists. Physicians typically attend to nearly 50 patients in a single morning, while essential medical equipment such as blood glucose monitors and insulin pumps remains severely inadequate. As Participant 11 (P11) noted, "The doctor who treated me worked very hard, but I still waited in line for three hours before entering the clinic." These findings underscore the urgent need for scalable, AI-assisted solutions that can support healthcare delivery and alleviate the workload in resource-limited areas.

### 3) Difficulties due to late diagnosis and delayed treatment

Doctor P40 said "If the disease can be detected early and intervened in time, patients will not suffer such severe diseases and will not experience so much pain. But the problem is that it is difficult for patients to detect the early symptoms of chronic diseases." P39 further explained that "When we try to treat, many patients' conditions are almost incurable. All we can do is to maintain the patient's life through insulin injections. In any case, I hope that patients can be diagnosed and treated as early as possible. If it is in Beijing or Shanghai, the number of patients who wait for their condition to worsen before coming for treatment will be reduced."

### 4) Insufficient follow-up and lifestyle intervention

The lack of and limited awareness of follow-up and lifestyle intervention appeared as the next obstacle. Many respondents, such as P2, P3, P5, P7, P11, and P14, went to the hospital for the treatment of T2DM complications, such as necrosis of the legs and feet and blurred vision. The treatment temporarily relieved their condition. However, they lacked awareness of subsequent intervention and follow-up after hospital discharge. For example, P14 said, "I came to the hospital because of blurred vision. The doctor said I had type 2 diabetes, but I didn't understand. The doctor prescribed me medicine and I took it. Now my vision has recovered. Why do I need to continue follow-up and medication?"

### 5) Self-management challenges

The interviews revealed limited self-monitoring abilities of many T2DM patients. Except for P21, P22, and P27, all other participants used blood glucose meters less frequently than doctor recommendation, and P33 had never used a blood glucose meter. Some respondents judged their blood glucose levels based on symptoms alone. In addition, the respondents generally lack scientific exercise guidance, and many equate exercise with work.

## III. SYSTEM DESIGN

Based on the needs and difficulties identified through participant interviews, we developed an AI-driven mobile application designed to help patients with type 2 diabetes mellitus (T2DM) communicate more effectively with their physicians and manage their health conditions. The application, named T2DM Health, is compatible with both Android and iOS platforms and integrates several core functions, including the interpretation of medical terminology for lay users, the explanation of treatment plans and diagnostic reports, patient condition tracking with automatic summary report generation, and personalized health guidance on diet and exercise (Figure 1)

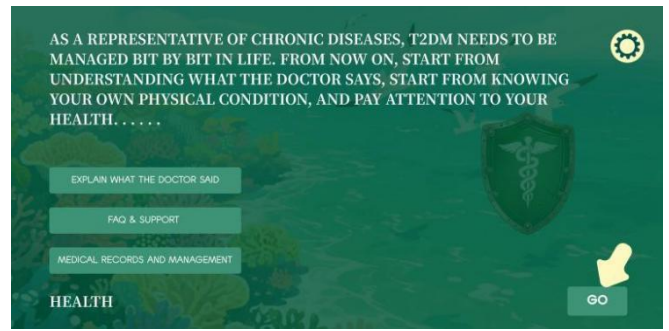

Fig. 1. The main interface of the system. The text in the figure introduces the three main functions of the system to users, including enhancing doctor-patient communication, answering questions, and assisting user management. Original text in Chinese is translated to English.

### A. Core Functionalities

Core functionalities of T2DM Health center around the entire lifecycle of doctor-patient interaction, from pre-visit preparation to ongoing support after discharge.

### 1) Pre-visit assessment

This function was designed to address the lack of follow-up identified during patient interviews. After a patient schedules an appointment, the application prompts them to provide disease-related information through an adaptive assessment module. Instead of relying on a fixed set of questions, the system dynamically evaluates the patient's existing knowledge and level of understanding during the initial interaction, then tailors subsequent questions to identify potential knowledge gaps. This adaptive approach can reduce patient fatigue caused by lengthy and rigid questionnaires while enabling more accurate assessments of health literacy and disease awareness. For example, the app may ask questions in a natural conversational form, such as "Have you measured your blood sugar?" or "What is your usual reading?"

Data from the pre-visit assessment will then be processed by GPT-4o to generate a detailed summary, including: (a) the patient's chief complaint and how it changes over time,

(b) specific concerns expressed by the patient, (c) key questions the patient asked, and (d) any significant emotional or behavioral patterns observed. This format would allow doctors to quickly understand the patient's situation within limited time of a clinical visit. The system also highlights key issues that require special attention, such as "the patient repeatedly expressed concerns about insulin side effects" and "poor blood sugar control, with potential medication compliance issues", or significant deviations from expected treatment path.

*2) Augmentation of clinical visits*

As the core innovation of our design, this module provides three sub-functions: real-time patient-physician conversation transcription, terminology explanation for patients, and intelligent medical record organization.

The real-time conversation transcription function is based on the Paraformer-MTL-v1 speech recognition engine provided by Alibaba Cloud, which is able to capture the content of patient-physician conversations with a high accuracy while effectively handling various Chinese dialects. The system automatically recognizes medical terms in the conversation and displays them in a sidebar that the patient can view during and after the visit. Users can click on any term in the list to get its definition and explanation sourced from a central medical knowledge graph. The explanations are also provided in a way that users can cross-reference the conversation report. This interactive approach allows patients to learn at their own pace, better improving their understanding of the conversation.( Figure 2)

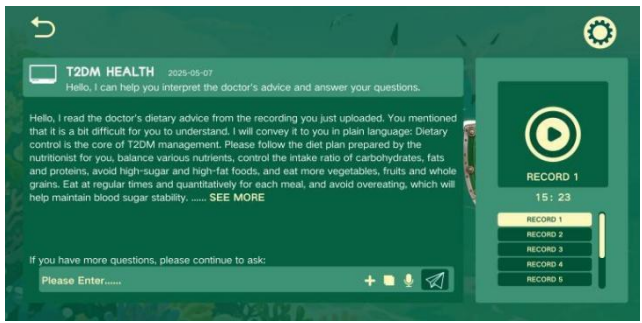

Fig. 2. Main interface for doctor-patient communication. The right side records the original data of the doctor- patient communication, and the left side provides explanatory communication, allowing users to ask questions in multiple forms. Original text in Chinese is translated to English.

The intelligent medical record organization feature is designed to provide clinicians with a structured, continuously updated, and easy-to-understand overview of each patient's medical course and health status. The function is achieved by integrating information from multiple sources, including doctor-patient conversation records, semi-structured assessment reports (including pre-visit assessment and health tracking reports, see Sections III.A.1 and III.A.3), and patient- reported data, to generate concise summaries, highlighting important problems and organizing information chronologically and thematically. This design enables doctors to efficiently review the patient's medical history, understand changes in their condition, and identify key issues before or during follow-up appointments, thereby improving clinical communication. These organized records are also the basis of monthly reports described in Section III.C.

*3) Post-visit support*

This function targets problems such as self-management difficulties and limited access to health information of the patients by providing medication reminders, health tracking, AI-driven questioning and answering (Q&A), etc.

The medication reminder system uses local push notifications (iOS uses User Notification Center, Android uses Notification Manager) to set personalized reminders based on the patient's medication habits and daily schedule. The system not only reminds patients to take medication on time, but also displays the name, dosage and purpose of the medication, such as "Now is the time to take metformin. Taking it after meals will help control blood sugar".

The health tracking function acts through daily conversational interactions. Similar to personalized pre-visit assessments in Section III.A.1, symptom tracking questions are made adaptive and insightful by adopting AI-driven semi- structured interviews. The system would proactively ask about the patient's conditions at the right time, such as "How did you sleep last night? Did you wake up at night?" in the morning, and "Have you measured your blood sugar after the meal? Do you feel uncomfortable?" after mealtime. Collected data are sorted and classified to form a continuous record of the patient's conditions.( Figure 3)

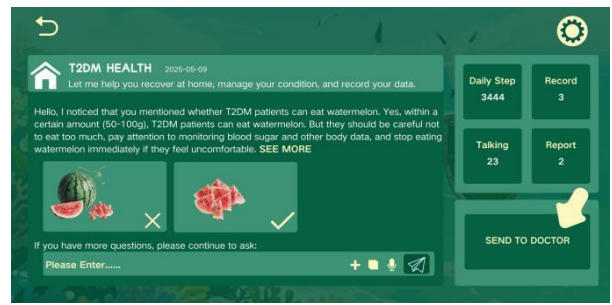

Fig. 3. The system supports users in daily life management and data recording. Original text in Chinese is translated to English.

The AI Q&A and educational content recommendation function based on GPT-4o combines the patient's personal characteristics (e.g., education level) with a medical knowledge base to provide 24-hour online consultation services. This function supports voice input and dialect recognition and pays special attention to the needs of rural patients. For example, when a patient asks, "Can I eat fruit when I take diabetes medication?", the system will provide personalized suggestions based on the patient's medication and blood sugar control conditions.

*4) Family collaboration*

This feature allows patients to authorize their dependents to view their health status while tightly controlling permission levels. Families can view the patient's medication status, blood glucose trends, and medical consultation history. The system also offers a "Care Mode" that automatically notifies authorized family members when a possible emergency is detected for a patient, such as failing to track blood glucose for several days in a row, having abnormally high blood sugar, or missing a medication. This design takes into account the situation of seniors who live alone and provides them with an extra layer of security.

The family collaboration function uses WebSocket for timely communication [21]. Permission management is based on the role-based access control model [22], allowing patients to set the permissions of different family members flexibly.

*B. System Architecture*

The mobile app system adopts a robust three-tier architecture, including a presentation layer, a logic layer, and a data layer, to ensure scalability, performance, and availability across different user groups.

*1) Presentation layer*

This layer contains multi-platform user interfaces and interaction modules. Considering the common demographic characteristics of T2DM patients, our T2DM Health app is especially optimized for the elderly, supporting multimodal inputs in both voice and text. The front-end uses React 18.3.1 and Tailwind CSS for a responsive design and consistent user experience across different devices types. App development used Kotlin and Jetpack Compose for Android systems and Swift and SwiftUI for iOS. A WebView browser engine was used for web-based functions. The user interface design of the app considers the characteristics of elder patients. For example, the buttons are enlarged by 20%, the default font is enlarged, the color contrast is improved by 30%, and the operation process is simplified to a maximum of three steps to complete any task. The recording function adopts a one-click design. Patients can start recording by simply touching the screen without complicated setup steps.

*2) Logical layer*

This layer serves as the core processing unit of the system and integrates multiple modules that drive the key functionalities described above. The AI engine supporting both the pre-visit assessment and the post-visit Q&A continuously evaluates patients' understanding and knowledge gaps, which are then used to generate explanations and recommend educational content tailored to individualized treatment plans. The AI engine and its underlying models are continuously refined by aggregating anonymized patient interaction data, including frequently asked questions and common knowledge patterns. The context-aware medical terminology explanation module leverages a term graph generated by the LightRAG framework and stored in Neo4j as a structured knowledge graph. This graph connects complex medical concepts to simplified, easy-to-understand explanations, alternative expressions, and disambiguation cues. The retrieval-augmented generation (RAG) pipeline, which powers the AI-driven Q&A and educational content recommendation features, adopts a hybrid retrieval strategy that combines precise, relation-aware term queries on the Neo4j knowledge graph (via the Cypher query language) with semantic search in a vector database to retrieve conceptually similar information. This design ensures the quality, relevance, and reliability of AI-generated responses. Finally, the treatment plan interpretation module employs GPT-4o to produce personalized explanations based on the patient's comprehension level. For instance, the system can not only restate a physician-recommended action plan but also enrich it with intuitive metaphors drawn from daily life to enhance understandability.

*3) Data layer*

The system adopts a hybrid data storage solution designed to maximize the advantages of different database systems. The PostgreSQL database deployed through the Supabase platform is responsible for storing structured data, including the patients' personal information, health records, and medication history. This choice was based on Supabase's excellent performance in real-time capabilities and security features, which is particularly suitable for processing sensitive medical data. The system implements end-to-end encryption to ensure data privacy, and all data transmission uses the TLS 1.3 protocol. The Neo4j graph database is capable of storing complex knowledge graph relationships. The medical terminology knowledge graph generated by LightRAG contains more than 500 medical terms along with common expressions and semantic relationships. The interactive history database uses the InfluxDB time series database, which is especially suitable for storing conversation records, symptom changes, and medication tracking data.

*C. System Learning and Feedback Mechanism*

The system automatically integrates multiple data streams to generate monthly patient health reports to provide clinicians with a comprehensive perspective about a patient's conditions, including blood sugar trend analysis, medication compliance assessment, and symptom frequency statistics. An important feature of the report is the incorporation of insights from the AI-driven semi-structured interviews about knowledge gaps and misconceptions of the patient. The report uses a clear visual design that allows doctors to quickly grasp the patient's overall status and progress over the past month.

Various technological methods were employed to obtain these analytical results. For instance, the pandas library in Python was utilized for robust data preprocessing, while scikit-learn was used for trend analysis and the detection of anomalies in patient health indicators. Recognizing the emotional states of patients during interactions is equally important; therefore, a pre-trained BERT model was applied for sentiment analysis of conversational data to highlight patients' anxiety, confusion, or satisfaction, providing physicians with valuable insights into the psychological state of patients throughout the entire process. In addition, the analysis involves identifying patterns in patient inquiries, frequently misunderstood concepts, and language differences found in AI-driven assessment and consultation. These combined insights form a critical feedback loop that is not only used to generate aforementioned patient reports for physician review, but also to iteratively optimize the AI pipeline, such as updating the medical knowledge graph with LightRAG and optimizing the prompting strategy for GPT-4o.

In summary, the proposed system design is capable of addressing the challenges revealed in initial interviews through a combination of multiple technical solutions. The system builds a bridge from clinical visits to daily health management, providing multifaceted health support for T2DM patients.

## IV. YSTEM EVALUATION

*A. Study Design*

Besides system performance metrics (e.g., speech recognition accuracy and latency, medical term recognition accuracy, and medical question answering correctness), we conducted a controlled experiment with follow-up qualitative interviews to assess the effectiveness and usability of our T2MD Health app. The primary outcome was patient performance in knowledge tests before and after using the app. Secondary outcomes contained system usability, quality ratings of AI-generated content by physicians, and patient self-management behaviors. This mixed method design helps answer the following evaluation questions: (1) How effective was the T2MD Health system in improving knowledge-heavy communication between patients and providers? (2) Whether and how could the system assist patient in disease self-management? (3) Was the system user-friendly and easy to use? What are some remaining defects of the system that can inform future work?

The T2DM knowledge test includes 23 multiple-choice and 4 true/false questions on the basic concepts of T2DM, and 1 long open-ended question about a patient's understanding of their conditions, treatment plans, and self-management actions. The open-ended question is scored by a physician. The total score of the knowledge test is 50.

We recruited 40 T2DM patients and 20 physicians online via RedNote in March 2025. Inclusion criteria were T2DM patients or doctors. Exclusion criteria were very severe T2DM or related malignant diseases. The 60 participants in the evaluation study were different from the 40 participants of the initial interviews. Participants went through the following experiment procedure:

- Step 1: All patients take the T2DM knowledge test. Patients were then divided into two groups, each with 20 patients, based on their initial test score to ensure that both groups had a similar average score in the knowledge test.

- Step 2: Each patient in the intervention group has a simulated clinical visit with a physician with the T2DM Health app open for conversation recording, whereas control patients have simulated clinical visits without the app.

- Step 3: After the visit, intervention patients are instructed to view the augmented vist summary generated by the T2DM Health app and interact with it.

- Step 4: Patients in both groups take the knowledge test again.

- Step 5: Intervention patients continue to have access to the app during the following 4 weeks. The control group do not.

- Step 6: Patients in both groups take the knowledge test for the third time.

- Step 7: Each patient has a simulated follow-up visit with their assigned physician. If a patient is in the intervention group, the physician receives the monthly health report generated by the T2DM Health app, and the visit is augmented by the app similar to step 2.

- Step 8: Patients in both groups take the knowledge test for the last time.

- Step 9: Physicians rate the quality of app-generated reports in terms of accuracy (40%), relevance (30%), readability (20%), and user-friendliness (10%). Physicians and patients in the intervention group evaluate the usability of the app and are interviewed about their experience with the app.

The study design was reviewed and approved by the Institutional Review Board of [anonymized] University. We acquired informed consent from all participants and provide the option of unconditional withdrawal from the study at any time. All data were deidentified before analysis.( Figure 4)

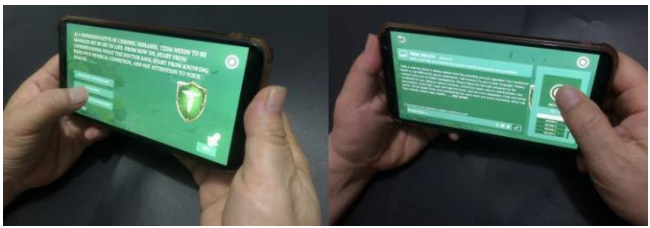

Fig. 4. User process record of using the system.

## B. Evaluation Results

In system performance testing, the T2MD Health system accurately recognized 96.2% of 500 standard Mandarin speech samples and 87.0% of 100 major dialect samples. The maximum delay from voice input to text display was 1.4 seconds, which meet real-time needs. The system accurately identified 92.2% of medical terms in the sample data. When tested with a sample of 200 medical questions, the system achieved an answer accuracy rate of 97.5%.

The 40 patient participants of the controlled experiment had an average age of 42.5 years (SD=13.7) and an average T2DM history of 5.7 years (SD=5.5). Half of the patients (n=20) lives in rural areas and 55% (n=22) were female. The 20 physician participants had an averge age of 46.0 years (SD=9.3) and a median work history of 14.0 years (Q1=7, Q3=26.25) as physicians, among which 55% (n=11) were female. All participants completed the full experiment.

Table 1 shows the summary statistics of knowledge test results at each time points. Test scores of the two groups were compared using the student's t-test or Mann-Whitney U test depending on data normality. Both groups had improved mean test scores over time. The difference between mean test scores of the two groups were not statistically significant before and after the initial visit. After 4 weeks, the difference became statistically significant, indicating that the intervention with our T2DM Health app was effective in promoting patient knowledge and understandings of T2DM.

TABLE I. MEAN (SD) KNOWLEDGE TEST SCORES AT THE 4 TIME POINTS.

|  | Intervention n=20 | Control n=20 | $P$ |
|---|---|---|---|
| Before initial visit | 32.8 (6.5) | 32.2 (4.9) | .76 |
| After initial visit | 38.3 (5.0) | 35.9 (4.5) | .12 |
| Before follow-up | 42.2 (3.6) | 36.4 (4.9) | <.001 |
| After follow-up | 45.6 (2.2) | 40.0 (4.3) | <.001 |

Qualitative interviews with intervention participants and secondary measures further revealed the following findings:

*1) The app enhanced patient-provider communication*
Almost all participants mentioned in qualitative interviews
That the T2DM Health app enhanced the effectiveness of patient-provider communication. Patient 11 said "The T2MD health system explained what prediabetes is in a way that I can understand, which is great." Patient 14 said "During the follow-up visit, the doctor directly viewed the data recorded in the system, and I didn't need to describe the information in words. My language expression ability is limited, and it is much more effective for the doctor to directly view the records than to listen to me." Physicians also praised the app. For example, doctor 5 said "During the first visit, I felt that this system helped explain the condition and the patient should be able to understand it. During the follow-up visit, I was already very familiar with the process and no longer needed to explain it to the patient. It was like communicating with a colleague who also understood it." Another perceived benefit of the app is its time-saving ability. Doctor 2 said "I think the biggest contribution of the T2MD Health system is time saving. This is also the place that can be further optimized in the next step. The actual situation of our hospital, department and even the country is that I need to see many patients, and the time I can allocate to each patient is very limited. I hope the system can help me convey as much information as possible in a limited time." Similarly, doctor 6 mentioned "There are many, many things, such as coronary atherosclerosis, the difference between T1DM and T2DM, that I no longer need to explain (because I know that the T2MDhealth system will explain), which saves a lot of time."

*2) Physician approval of app-generated content*

As shown in Table 2, the content generated by the T2DM Health app received high scores in both facet-wise ratings and overall ratings. None of the physicians reported safety concern about app-generated content.

TABLE II. PHYSICIAN RATINGS OF APP-GENERATED CONTENT OUT OF 30.

| Facet | Mean (SD) |
|---|---|
| T2DM basic knowledge | 27.6 (2.6) |
| T2DM treatment information | 25.8 (2.4) |
| Health tracking | 26.8 (2.4) |
| Lifestyle management | 26.7 (2.0) |
| Psychological support | 24.3 (1.8) |
| Risk prevention | 25.7 (2.1) |
| Overall | 26.0 (1.1) |

information flow and that the app is easy to use with a low learning curve. As shown in Table 3, patient assessments using the system usability scale (SUS) further supported this finding.

TABLE III. PATIENT SUS RATINGS. EACH ITEM IS ON A SCALE OF 1 TO 5.

| SUS Item | Mean (SD) |
|---|---|
| I think I would use this system frequently | 4.8 (0.4) |
| I found the system unnecessarily complex | 1.2 (0.4) |
| I thought the system was easy to use | 4.5 (0.5) |
| I would need technical support to use this system | 1.2 (0.5) |
| I found the various functions well integrated | 4.6 (0.5) |
| I thought there was too much inconsistency | 1.0 (0.2) |
| Most people would learn to use this system quickly | 4.6 (0.5) |
| I found the system very cumbersome to use | 1.2 (0.6) |
| I felt confident when using this system | 4.6 (0.4) |
| Overall, I would not recommend this system | 1.1 (0.3) |

*3) Remaining defects and improvement suggestions*

Open-ended responses regarding improvement suggestions indicated that the T2MD Health app still had defects in data processing and analysis. First, the system's recognition accuracy for some suburban accents is not high enough, and the dialect library needs to be expanded. For example, patient 9 mentioned "Maybe my accent is too strong. The content I input by voice is inconsistent with the text recognized by the system." Second, the app did not dive deep into complication complexities, and physician review is needed. Doctor 7 mentioned "Many diseases are related to T2DM. The current logical framework can be richer, because the world and the human body are complex." Third, physician mentioned that the system focused more on the accuracy of basic knowledge rather than the understanding of the specific symptoms of the patients. Doctor 11 mentioned: "(It) generates a lot of correct nonsense. For example, patients ask whether they should follow up at a certain time. Ideally, the system should make judgments based on the information provided by the patient. It should ask the patient to provide more information if the information is insufficient. However, the records I saw showed that it generated a lot of general knowledge about when T2DM patients should follow up. This is indeed correct information, but you know, it is not relevant enough to the patient's question."

*4) The app alleviated challenges faced by rural patients*

Most (7 out of 9) rural patients in the intervention group expressed that the app was helpful in addressing the urban-rural gap in the access to medical consultation services. For example, patient 2 said "It provides a platform for me to get medical resources without going to the city."

*5) The app's usability was high*

Qualitative interviews with physicians revealed that the app could help reduce physician workload by streamline the

## V. CONCLUSION & DISCUSSION

This study revealed the challenges faced by T2MD patients in patient–provider communication and self-management, which guided the design of an AI-driven, multifunctional mobile app that enhances communication between patients and doctors, supports patient self-management, and helps bridge the gap between urban and rural medical care. Compared with previous systems that focus primarily on clinic-based or one-way digital monitoring [23, 24], our system offers a bidirectional, multimodal interaction framework that enables real-time dialect interpretation and continuous AI-based feedback. Prior studies have rarely integrated dialect-specific natural language processing or context-aware patient education, both of which were considered and addressed in our approach.

The core contribution of this work lies in its formative, user-centered study of a system tailored to Chinese T2DM patients. Nevertheless, this study has several limitations. First, as a pilot experiment, it involved a relatively small number of participants. All participants were recruited through a single online platform (RedNote), which may have resulted in a sample with higher digital health literacy than the general T2DM population in China. To further evaluate the system's effectiveness and safety, future research should conduct larger-scale trials with more diverse participants, a longer study period (for example, 12 months), and multiple study groups for comparison, such as a control group and an intervention group using another health education app.

First, as a pilot study to demonstrate the feasibility and usability of the app, our evaluation experiment involved a relatively small number of participants. Meanwhile, participants were recruited through a single source (RedNote) online, potentially leading to a study population that possesses higher digital health literacy than the general T2DM patient population in China. To further evaluate the effectiveness and safety of our proposed system design, a larger-scale trial involving participants from more diverse backgrounds, a longer study window (e.g., 12 months), and multiple study groups for better comparison (e.g., a blank control group and an intervention group with another health education app) will be necessary.

Second, the limitations identified through qualitative interviews provide directions for improvement. One is dialect recognition accuracy, given the diversity of Chinese dialects. The relevance of AI-generated responses to patient questions could also be enhanced through techniques such as question rewriting, thresholded retrieval, and context engineering. As AI models for speech transcription and user interaction continue to evolve, we expect better overall user experience in future iterations of the app.

Third, additional data modalities could be integrated to improve analysis, such as blood glucose readings from wearable devices. A physician-in-the-loop mechanism could also be established to further enhance consultation quality, particularly for complex cases with comorbidities. Moreover, low-bandwidth technical solutions, including edge-based AI, could better support areas with limited Internet access.

Beyond these technical considerations, our findings suggest that the value of AI in healthcare lies not in automation but in mediation, bridging the cognitive, linguistic, and relational gaps that have long constrained chronic disease care. One of the key implications of this study is the shift from passive patient education to active, patient-generated understanding. Through conversational interaction, patients not only receive information but co-construct meaning with the AI system, transforming everyday dialogue into a process of continuous knowledge reinforcement. This perspective points to a future in which digital health systems act as *cognitive scaffolds*, fostering lasting comprehension, confidence, and self-efficacy in disease management [27, 28].

The deployment of T2MD Health also raises questions about how AI should participate in clinical communication. Our findings indicate that AI is not a replacement for human explanation but an intermediary that strengthens trust and understanding between patients and physicians. By transcribing, explaining, and contextualizing medical information, the system helps patients engage in consultations with greater clarity and confidence, potentially reshaping traditional power dynamics in doctor–patient interactions.

As such systems begin to influence how medical understanding is produced and shared, issues of governance and accountability become integral to design. Future versions of T2MD Health should ensure transparency, privacy, and informed consent, particularly when patient dialogues are stored or analyzed. Moreover, as large language models continue to expand, sustainability and accessibility remain critical. Lightweight or edge-based implementations may offer a more responsible path forward for low-resource healthcare environments while guiding future principles for equitable, human-centered AI in health.